\documentclass[aip,graphicx]{revtex4-1}
\usepackage [latin1]{inputenc}
\draft 
\usepackage {amssymb}
\usepackage {amsmath}
\begin{document}

\title{Symmetry evolution for the imperfect fluid under perturbations} 

\author{Alcides Garat}
\affiliation{1. Former Professor at Universidad de la Rep\'{u}blica, Av. 18 de Julio 1824-1850, 11200 Montevideo, Uruguay.}
\email[]{garat.alcides@gmail.com}
\date{\today}

\begin{abstract}
Ever since a new symmetry was found for the imperfect fluid with vorticity the question of the effect of perturbations on the symmetry itself has been raised. This new symmetry arose when realizing that local four-velocity gauge-like transformations would render the left hand side of the Einstein equations invariant. Because the metric tensor would be invariant under this new kind of local transformations. Then the point was raised about the invariance of such a kind of transformations of the stress-energy tensor on the right hand side of the Einstein equations in curved four-dimensional Lorentz spacetimes. It was verified that these invariances do not work with plain perfect fluid but they do work for imperfect fluids. The imperfect fluid stress-energy tensor will be invariant under local four-velocity gauge-like transformations when additional transformations are introduced for several variables included in the stress-energy tensor itself. This local invariance was also the criteria introduced in order to present a new stress-energy tensor for vorticity as well. New tetrads are at the core of the realization of the existence of this new symmetry because it is through these new tetrads that this new symmetry is realized. It is through the local transformation of these new tetrad vectors that we can prove that the metric tensor is invariant. This new kind of symmetry has its origins in a similar tetrad formulation as to the Einstein-Maxwell spacetimes formalism presented in previous manuscripts. In this manuscript we will introduce local perturbations by external agents to the relevant objects in the imperfect fluid geometry. We will demonstrate a theorem that proves that the symmetries under four-velocity gauge-like transformations will be instantaneously broken but at the same time transformed into new symmetries. Because the local orthogonal planes determined by these new tetrads, which happen to be the local planes of symmetry will tilt under local perturbations. There will be a symmetry evolution under perturbations.
\end{abstract}



\maketitle 

\section{Introduction}
\label{intro}

In manuscripts \cite{AGNS,AGNSHC} it was proved that the existence of new tetrads in curved four-dimensional Lorentz spacetimes  with signature (-+++) enabled the proof about a new symmetry of spacetime in the presence of imperfect fluids with vorticity. These tetrads have remarkable properties that create this new mathematical scenario where new relevant geometrical features are visualized with simplicity. These new tetrads are built with two structure components, the skeletons and the gauge vectors. The timelike-spacelike local plane, we call plane one. The orthogonal local spacelike-spacelike plane is plane two.

However in our present case we will start our analysis with an imperfect fluid where the stress-energy tensor of a source can be described by the following equation \cite{EG,AC,F,JO},

\begin{equation}
T^{imp}_{\mu\nu}= (\rho + p)\:u_{\mu}\:u_{\nu} + p\:g_{\mu\nu} + (q_{\mu}\:u_{\nu} + q_{\nu}\:u_{\mu}) + \tau_{\mu\nu}\ ,\label{SETIMPNP}
\end{equation}

where the viscous stress-energy tensor $\tau_{\mu\nu}$ is given by \cite{JO},

\begin{equation}
\tau_{\mu\nu}= -\eta\:\left(u_{\mu;\nu} + u_{\nu;\mu} + u_{\mu}\:u^{\alpha}\:u_{\nu;\alpha} + u_{\nu}\:u^{\alpha}\:u_{\mu;\alpha}\right) - (\zeta-\frac{2}{3}\:\eta)\:u^{\alpha}_{\:\:;\alpha}\:(g_{\mu\nu} + u_{\mu}\:u_{\nu}) \ ,\label{SETIMPVISCNP}
\end{equation}

where $\rho$ is the energy-density of the fluid, $p$ the isotropic pressure and where $q_{\mu}$ is the heat flux relative to $u^{\mu}$ its four-velocity field, $g_{\mu\nu}$ is the metric tensor. The parameter $\eta$ is the coefficient of shear viscosity and the parameter $\zeta$ is the coefficient of bulk viscosity. If in addition this fluid has vorticity $\omega_{\mu\nu}$, then we can build the new tetrads for this particular case following the method developed in reference \cite{A,AGE,ROMP,SING,ATGU} and explained from different angles in references \cite{IWCP,LomCon,AEO,ACD,ENV,MW}. These new tetrads introduced and described in detail in manuscript \cite{AGNS} will manifestly and covariantly diagonalize the stress-energy tensor for the perfect fluid terms $T^{p}_{\mu\nu}= (\rho + p)\:u_{\mu}\:u_{\nu} + p\:g_{\mu\nu}$ at every spacetime event. However in this manuscript we are interested in the study of the local symmetries when an external agent perturbs the source of relevant fields. Source of gravitational field, a source with a stress-energy tensor given by equations (\ref{SETIMPNP}) and (\ref{SETIMPVISCNP}). For perturbations to imperfect fluids see references \cite{SSAK,EC,LMRB} and references therein.

In order to find the new tetrads for the new perturbation problem we proceed then to introduce the fluid perturbed extremal field or the perturbed velocity curl-extremal field through the local duality transformation given by,

\begin{equation}
\overline{\xi}_{\mu\nu} = \cos\overline{\alpha} \:\: \overline{u}_{[\mu;\nu]} - \sin\overline{\alpha} \:\: \ast \overline{u}_{[\mu;\nu]} ,\label{vef}
\end{equation}

where $\ast \overline{u}_{[\mu;\nu]} = {1 \over 2}\:\epsilon_{\mu\nu\sigma\tau}\:\overline{g}^{\sigma\rho}\:\overline{g}^{\tau\lambda}\:\overline{u}_{[\rho;\lambda]}$ is the dual tensor of $\overline{u}_{[\mu;\nu]}$ and the local complexion $\overline{\alpha}$ is defined through the condition

\begin{equation}
\overline{\xi}_{\mu\nu}\:\ast \overline{\xi}^{\mu\nu} = 0 \ .\label{fcond}
\end{equation}

The symbol $;$ stands for covariant derivative with respect to the metric tensor $\overline{g}_{\mu\nu}$ and the object $\epsilon_{\mu\nu\sigma\tau}$ is the alternating tensor, see reference \cite{A}. The objects with tildes represent the perturbed objects. For example $\overline{u}^{\mu}=(u^{\mu}+\varepsilon\:\delta u^{\mu})/\sqrt{|(u^{\rho}+\varepsilon\:\delta u^{\rho})\:\overline{g}_{\rho\lambda}\:(u^{\lambda}+\varepsilon\:\delta u^{\lambda})|}$ where $\delta u^{\mu}$ is a local perturbation to the local unit four-velocity $u^{\lambda}$ with a suitable perturbation parameter $\varepsilon$ and $\overline{u}^{\mu}\:\overline{g}_{\mu\nu}\:\overline{u}^{\nu}=-1$. For the gravitational field as another example $\overline{g}_{\mu\nu}=g_{\mu\nu}+\varepsilon\:\delta g_{\mu\nu}$ where $\delta g_{\mu\nu}$ is a local perturbation to the local metric tensor with the same suitable perturbation parameter $\varepsilon$. All these perturbations are of a physical nature and not the result of a coordinate transformation. Similar for all the other variables in the physical setup. It must be stressed that even if the notation and the paper structure are similar to reference \cite{AGNS} the content is different because unlike manuscript \cite{AGNS} in this manuscript we are studying imperfect fluids under perturbations. We are trying to prove the dynamic nature of local symmetries and this proof requires a similar paper structure and notation with respect to reference \cite{AGNS} but the content is substantially different. Following exactly the same construction steps as in sections I and II in manuscript \cite{AGNS} we would find the following results, this time for the perturbed quantities. The complexion, which is a local scalar, can then be found by plugging equation (\ref{vef}) into equation (\ref{fcond}) as,

\begin{equation}
\tan(2\overline{\alpha}) = - \left( \overline{u}_{[\mu;\nu]}\:\overline{g}^{\sigma\mu}\:\overline{g}^{\tau\nu}\:\ast \overline{u}_{[\sigma;\tau]}\right) \:/\: \left(\overline{u}_{[\lambda;\rho]}\:\overline{g}^{\lambda\alpha}\:\overline{g}^{\rho\beta}\: \overline{u}_{[\alpha;\beta]}\right) \ .\label{complexion}
\end{equation}

After introducing the new perturbed velocity curl-extremal field we proceed to write the four perturbed orthogonal vectors that will become an intermediate step in constructing the tetrad that diagonalizes the perturbed perfect fluid stress-energy tensor terms as $\overline{T}^{p}_{\mu\nu}= (\overline{\rho} + \overline{p})\:\overline{u}_{\mu}\:\overline{u}_{\nu} + \overline{p}\:\overline{g}_{\mu\nu}$ (see further into the manuscript in equation (\ref{SETIMP}) and a detailed analysis in reference \cite{AGNS}),

\begin{eqnarray}
\overline{V}_{(1)}^{\alpha} &=& \overline{\xi}^{\alpha\lambda}\:\overline{\xi}_{\rho\lambda}\:X^{\rho}
\label{V1}\\
\overline{V}_{(2)}^{\alpha} &=& \overline{\xi}^{\alpha\lambda} \: X_{\lambda}
\label{V2}\\
\overline{V}_{(3)}^{\alpha} &=& \ast \overline{\xi}^{\alpha\lambda} \: Y_{\lambda}
\label{V3}\\
\overline{V}_{(4)}^{\alpha} &=& \ast \overline{\xi}^{\alpha\lambda}\: \ast \overline{\xi}_{\rho\lambda}
\:Y^{\rho}\ ,\label{V4}
\end{eqnarray}

In order to prove the orthogonality of the tetrad (\ref{V1}-\ref{V4}) it is necessary to use the identity

\begin{eqnarray}
A_{\mu\alpha}\:B^{\nu\alpha} -
\ast B_{\mu\alpha}\: \ast A^{\nu\alpha} &=& \frac{1}{2}
\: \delta_{\mu}^{\:\:\:\nu}\: A_{\alpha\beta}\:B^{\alpha\beta}  \ .\label{i1}
\end{eqnarray}

which is valid for every pair of antisymmetric tensors in a four-dimensional Lorentzian spacetime \cite{MW} for the case $A_{\mu\alpha} = \overline{\xi}_{\mu\alpha}$ and $B^{\nu\alpha} = \overline{\xi}^{\nu\alpha}$ we would obtain,

\begin{eqnarray}
\overline{\xi}_{\mu\alpha}\:\overline{\xi}^{\nu\alpha} - \ast \overline{\xi}_{\mu\alpha}\: \ast \overline{\xi}^{\nu\alpha} &=& \frac{1}{2}
\: \delta_{\mu}^{\:\:\:\nu}\:\overline{Q}\ ,\label{i2}
\end{eqnarray}

where $\overline{Q}=\overline{\xi}_{\mu\nu}\:\overline{\xi}^{\mu\nu}$ is assumed not to be zero. When (\ref{i1}) is applied to the case $A_{\mu\alpha} = \overline{\xi}_{\mu\alpha}$ and $B^{\nu\alpha} = \ast \overline{\xi}^{\nu\alpha}$ we obtain an equivalent condition to (\ref{fcond}),

\begin{equation}
\overline{\xi}_{\mu\rho}\:\ast\overline{\xi}^{\mu\lambda} = 0 \ .\label{scond}
\end{equation}

The vector fields $X^{\alpha}$ and $Y^{\alpha}$ in equations (\ref{V1}-\ref{V4}) are the gauge vectors and we are free to choose them as long as the four vector fields (\ref{V1}-\ref{V4}) do not become trivial. The tetrad vectors have two essential components. For instance in vector $\overline{V}_{(1)}^{\alpha}$ there are two main structures. First, the skeleton, in this case $\overline{\xi}^{\alpha\lambda}\:\overline{\xi}_{\rho\lambda}$, and second, the gauge vector $X^{\rho}$. In vector $\overline{V}_{(3)}^{\alpha}$ the skeleton is $\ast \overline{\xi}^{\alpha\lambda}$, and the gauge vector $Y_{\lambda}$. It is clear that if our choice for these gauge vector fields is $X^{\alpha} = Y^{\alpha} = \overline{u}^{\alpha}$, then the following orthogonality relations will hold,

\begin{eqnarray}
\lefteqn{ \overline{g}_{\rho\mu}\:\overline{u}^{\rho}\:\overline{V}_{(2)}^{\mu} = \overline{g}_{\rho\mu}\:\overline{u}^{\rho}\:\overline{\xi}^{\mu\lambda}\:\overline{u}_{\lambda} = 0 \label{ortho1} } \\
&&\overline{g}_{\rho\mu}\:\overline{u}^{\rho}\:\overline{V}_{(3)}^{\mu} = \overline{g}_{\rho\mu}\:\overline{u}^{\rho}\:\ast\overline{\xi}^{\mu\lambda}\:\overline{u}_{\lambda} = 0 \ ,\label{ortho2}
\end{eqnarray}

because of the antisymmetry of the perturbed velocity curl-extremal field $\overline{\xi}_{\mu\nu}$. Then, at the points in spacetime where the set of four vectors (\ref{V1}-\ref{V4}) is not trivial, we can normalize,

\begin{eqnarray}
\hat{\overline{U}}^{\alpha} &=& \overline{\xi}^{\alpha\lambda}\:\overline{\xi}_{\rho\lambda}\:\overline{u}^{\rho} \:
/ \: (\: \sqrt{-\overline{Q}/2} \: \sqrt{\overline{u}_{\mu} \ \overline{\xi}^{\mu\sigma} \
\overline{\xi}_{\nu\sigma} \ \overline{u}^{\nu}}\:) \label{Uw}\\
\hat{\overline{V}}^{\alpha} &=& \overline{\xi}^{\alpha\lambda}\:\overline{u}_{\lambda} \:
/ \: (\:\sqrt{\overline{u}_{\mu} \ \overline{\xi}^{\mu\sigma} \
\overline{\xi}_{\nu\sigma} \ \overline{u}^{\nu}}\:) \label{Vw}\\
\hat{\overline{Z}}^{\alpha} &=& \ast \overline{\xi}^{\alpha\lambda} \:  \overline{u}_{\lambda} \:
/ \: (\:\sqrt{\overline{u}_{\mu}  \ast \overline{\xi}^{\mu\sigma}
\ast \overline{\xi}_{\nu\sigma}   \overline{u}^{\nu}}\:)
\label{Zw}\\
\hat{\overline{W}}^{\alpha} &=& \ast \overline{\xi}^{\alpha\lambda}\: \ast \overline{\xi}_{\rho\lambda}
\: \overline{u}^{\rho} \: / \: (\:\sqrt{-\overline{Q}/2} \: \sqrt{ \overline{u}_{\mu}
\ast \overline{\xi}^{\mu\sigma} \ast \overline{\xi}_{\nu\sigma}  \overline{u}^{\nu}}\:) \ .
\label{Ww}
\end{eqnarray}

In analogy with the electromagnetic case and without tampering with anything fundamental we assume for simplicity that $\overline{u}_{\mu} \ \overline{\xi}^{\mu\sigma} \ \overline{\xi}_{\nu\sigma} \ \overline{u}^{\nu} > 0$, $\:\:\:\overline{u}_{\mu}
\ast \overline{\xi}^{\mu\sigma} \ast \overline{\xi}_{\nu\sigma}  \overline{u}^{\nu} > 0$ and $-\overline{Q} > 0$. We also assume that $\hat{\overline{U}}^{\alpha}\:\hat{\overline{U}}_{\alpha}=-1$. Using the tetrad vectors (\ref{Uw}-\ref{Ww}) and the method developed in manuscript \cite{A} for the antisymmetric electromagnetic field, we can express the four-velocity curl in its maximal simplest form,

\begin{equation}
\overline{u}_{[\mu;\nu]} = -2\:\sqrt{-\overline{Q}/2}\:\:\cos\overline{\alpha}\:\:\hat{\overline{U}}_{[\alpha}\:\hat{\overline{V}}_{\beta]} +
2\:\sqrt{-\overline{Q}/2}\:\:\sin\overline{\alpha}\:\:\hat{\overline{Z}}_{[\alpha}\:\hat{\overline{W}}_{\beta]}\ .\label{VC}
\end{equation}

The metric tensor will be written as,

\begin{equation}
\overline{g}_{\alpha\beta} = -\hat{\overline{U}}_{\alpha}\:\hat{\overline{U}}_{\beta} + \hat{\overline{V}}_{\alpha}\:\hat{\overline{V}}_{\beta} +
\hat{\overline{Z}}_{\alpha}\:\hat{\overline{Z}}_{\beta} + \hat{\overline{W}}_{\alpha}\:\hat{\overline{W}}_{\beta}\ .\label{MT}
\end{equation}

The pair of vectors ($\hat{\overline{U}}^{\alpha}, \hat{\overline{V}}^{\alpha}$) span the local plane one. The vectors ($\hat{\overline{Z}}^{\alpha}, \hat{\overline{W}}^{\alpha}$) span the local orthogonal plane two. The gauge-like Abelian group of transformations of the four-velocity $X^{\alpha} = Y^{\alpha} = \overline{u}^{\alpha} \rightarrow X^{\alpha} = Y^{\alpha} = \overline{u}^{\alpha}+\Lambda^{\alpha}$ induces several kind of local tetrad transformations either in the local plane one or the local orthogonal plane two. We use the notation $\Lambda^{\alpha}=\Lambda_{,\beta}\:\overline{g}^{\beta\alpha}$ for $\Lambda$ a local scalar. In the local plane one the vectors that span this plane undergo local tetrad transformations under the local group LB1 leaving these vectors inside the original local plane. The group LB1 is composed by $SO(1,1)$ plus two discrete transformations two by two. One of them is the full inversion or minus the identity two by two. The other discrete transformation is a reflection given by $\Lambda^{o}_{\:\:o} = 0$, $\Lambda^{o}_{\:\:1} = 1$, $\Lambda^{1}_{\:\:o} = 1$,  $\Lambda^{1}_{\:\:1} = 0$. LB2 for the local tetrad transformations in plane two is just $SO(2)$. All these possible cases have been discussed in detail in manuscript \cite{A}. The metric tensor will remain invariant under LB1 and LB2 as proven in manuscript \cite{A} for the electromagnetic field. Under all these transformations the four-velocity curl $\overline{u}_{[\mu;\nu]}$ and the extremal field $\overline{\xi}_{\mu\nu}$ will also remain invariant. We would like to know about the Einstein equations under this ``four-velocity gauge transformation''. The Einstein equations are given by,

\begin{eqnarray}
\overline{R}_{\mu\nu} - \frac{1}{2}\:\overline{g}_{\mu\nu}\:\overline{R} &=& \overline{T}_{\mu\nu}= (\overline{\rho} + \overline{p})\:\overline{u}_{\mu}\:\overline{u}_{\nu} + \overline{p}\:\overline{g}_{\mu\nu}\ , \label{EFE}
\end{eqnarray}

as we notice from equation $\overline{T}_{\mu\nu}= (\overline{\rho} + \overline{p})\:\overline{u}_{\mu}\:\overline{u}_{\nu} + \overline{p}\:\overline{g}_{\mu\nu}$. Under the transformation $\overline{u}^{\alpha} \rightarrow \overline{u}^{\alpha}+\Lambda^{\alpha}$ we see that the right hand side will change because $\overline{u}_{\mu}$ will change. The metric tensor remains invariant as already proved, therefore the left hand side of equation (\ref{EFE}) and the metric tensor on the second term of $\overline{T}_{\mu\nu}$ will remain invariant. We will be able to write the transformed new stress-energy tensor as,

\begin{equation}
\overline{T}^{new}_{\mu\nu}= (\overline{\rho} + \overline{p})\:(\overline{u}_{\mu}+\Lambda_{\mu})\:(\overline{u}_{\nu}+\Lambda_{\nu}) + \overline{p}\:\overline{g}_{\mu\nu}\ .\label{SETMOD}
\end{equation}

Then the perfect fluid stress-energy tensor will not be invariant under gauge-like four-velocity transformations on its own for the perturbed formulation either. In order to make it invariant we would have to add to the right hand side of the Einstein equations in (\ref{EFE}) different kinds of terms including heat flow currents $(\overline{\rho} + \overline{p})\:(\overline{u}_{\mu}\:\overline{q}_{\nu} + \overline{u}_{\nu}\:\overline{q}_{\mu})$ where $\overline{u}^{\mu}\:\overline{q}_{\mu}=0$ is satisfied and terms with viscous stresses in the fluid $\overline{\tau}_{\mu\nu}$ where $\overline{u}^{\mu}\:\overline{\tau}_{\mu\nu}=0$ is satisfied as well, see references \cite{HS,CW,RW} and section \ref{invariance}. This kind of fluid will be ideal no longer, however we are interested in the invariance geometrical properties of the perturbed Einstein equations under the gauge-like four-velocity transformations.

\section{Perturbed imperfect fluid stress-energy tensor symmetry}
\label{invariance}

Let us consider a perturbed imperfect fluid stress-energy tensor of the kind \cite{JO},

\begin{equation}
\overline{T}^{imp}_{\mu\nu}= (\overline{\rho} + \overline{p})\:\overline{u}_{\mu}\:\overline{u}_{\nu} + \overline{p}\:\overline{g}_{\mu\nu} + (\overline{q}_{\mu}\:\overline{u}_{\nu} + \overline{q}_{\nu}\:\overline{u}_{\mu}) + \overline{\tau}_{\mu\nu}\ ,\label{SETIMP}
\end{equation}

where $\overline{q}_{\mu}$ is the perturbed heat flux relative to the unit velocity $\overline{u}_{\mu}$ and the perturbed viscous stress-energy tensor $\overline{\tau}_{\mu\nu}$ is given by \cite{JO},

\begin{equation}
\overline{\tau}_{\mu\nu}= -\eta\:\left(\overline{u}_{\mu;\nu} + \overline{u}_{\nu;\mu} + \overline{u}_{\mu}\:\overline{u}^{\alpha}\:\overline{u}_{\nu;\alpha} + \overline{u}_{\nu}\:\overline{u}^{\alpha}\:\overline{u}_{\mu;\alpha}\right) - (\zeta-\frac{2}{3}\:\eta)\:\overline{u}^{\alpha}_{\:\:;\alpha}\:(\overline{g}_{\mu\nu} + \overline{u}_{\mu}\:\overline{u}_{\nu}) \ .\label{SETIMPVISC}
\end{equation}

The parameter $\eta$ is the coefficient of shear viscosity and the parameter $\zeta$ is the coefficient of bulk viscosity. The perturbed velocity will have to be normalized as in section \ref{intro}. Under the four-velocity gauge-like transformation $\overline{u}^{\alpha} \rightarrow \overline{u}^{\alpha}+\Lambda^{\alpha}$ the stress-energy tensor (\ref{SETIMP}) will change and will not be invariant if this is the only transformation that we carry out. Let us remind ourselves that the perturbed Einstein equations are given now by,

\begin{eqnarray}
\overline{R}_{\mu\nu} - \frac{1}{2}\:\overline{g}_{\mu\nu}\:\overline{R} &=& \overline{T}^{imp}_{\mu\nu} \ . \label{EFEIMP}
\end{eqnarray}

The left hand side of equations (\ref{EFEIMP}) will remain invariant since it has been proven already in section \ref{intro} and reference \cite{A} that the metric tensor is explicitly invariant under $\overline{u}^{\alpha} \rightarrow \overline{u}^{\alpha}+\Lambda^{\alpha}$. The covariant derivatives, meaning the Christoffel symbols will also be invariant for the same reason. Therefore by using this symmetry argument we realize that the right hand side of equations (\ref{EFEIMP}) will also have to be invariant. The transformations $\overline{u}^{\alpha} \rightarrow \overline{u}^{\alpha}+\Lambda^{\alpha}$ on the right hand side of equations (\ref{EFEIMP}) will not leave the tensor $\overline{T}^{imp}_{\mu\nu}$  invariant if we apply only this kind of four-velocity gauge-like transformation. In order to make it invariant we need to introduce the following additional local transformations,

\begin{eqnarray}
\overline{\rho} &\rightarrow& \overline{\rho} + \widetilde{\rho} \label{rhotransf}\\
\overline{p} &\rightarrow& \overline{p} + \widetilde{p} \label{ptransf}\\
\overline{q}^{\mu} &\rightarrow& \overline{q}^{\mu} + \widetilde{q}^{\mu} \label{heattransf}\\
\overline{\tau}^{\mu\nu}(\overline{u}) &\rightarrow& \overline{\tau}^{\mu\nu}(\overline{u}+\Lambda) + \widetilde{\tau}^{\mu\nu} \ . \label{viscoustransf}
\end{eqnarray}

By $\overline{\tau}^{\mu\nu}(\overline{u})$ we mean the equation (\ref{SETIMPVISC}) while by $\overline{\tau}^{\mu\nu}(\overline{u}+\Lambda)$ we mean equation (\ref{SETIMPVISC}) under the transformation $\overline{u}^{\alpha} \rightarrow \overline{u}^{\alpha}+\Lambda^{\alpha}$. This notation will shorten the detailed writing of long expressions with many terms. There is also a state equation $\overline{p}(\overline{\rho})$ and there is also another one $\widetilde{p}(\widetilde{\rho})$. There is then only one local scalar independent variable that we consider to be $\widetilde{\rho}$. After imposing invariance of the perturbed stress-energy tensor $\overline{T}^{imp}_{\mu\nu}$ we will obtain ten equations for the fifteen variables $\widetilde{\rho}$, $\widetilde{q}^{\mu}$ and $\widetilde{\tau}^{\mu\nu}$. Additionally we also impose that $\overline{q}^{\alpha}\:\overline{u}_{\alpha}=0$ and $\overline{\tau}^{\mu\nu}\:\overline{u}_{\nu}=0$ which are five more equations.
The extension of these last two equations to the case under the transformation $\overline{u}^{\alpha} \rightarrow \overline{u}^{\alpha}+\Lambda^{\alpha}$ we find $(\overline{q}^{\mu} + \widetilde{q}^{\mu})\:(\overline{u}_{\mu}+\Lambda_{\mu})=0$ and $(\overline{\tau}^{\mu\nu}(\overline{u}+\Lambda) + \widetilde{\tau}^{\mu\nu})\:(\overline{u}_{\mu}+\Lambda_{\mu})=0$. From these last two ``transverse'' equations plus the original ones $\overline{q}^{\alpha}\:\overline{u}_{\alpha}=0$ and $\overline{\tau}^{\mu\nu}\:\overline{u}_{\nu}=0$ we obtain the following equations,

\begin{eqnarray}
\overline{q}^{\mu}\:\Lambda_{\mu} + \widetilde{q}^{\mu}\:(\overline{u}_{\mu}+\Lambda_{\mu})&=&0 \label{conditionsheat}\\
\overline{\tau}^{\mu\nu}(\overline{u})\:\Lambda_{\nu} + \overline{\tau}^{\mu\nu}(\Lambda)\:(\overline{u}_{\nu}+\Lambda_{\nu}) + \widetilde{\tau}^{\mu\nu}\:(\overline{u}_{\nu}+\Lambda_{\nu})&=&0 \ . \label{conditionsvisc}
\end{eqnarray}

By $\overline{\tau}^{\mu\nu}(\Lambda)$ we mean all the terms of the style $\overline{u}_{\mu}\:\Lambda_{\nu;\alpha}\:\overline{u}^{\alpha}+\overline{u}_{\nu}\:\Lambda_{\mu;\alpha}\:\overline{u}^{\alpha}$ or $\Lambda_{\mu}\:\overline{u}_{\nu;\alpha}\:\Lambda^{\alpha}+\Lambda_{\nu}\:\overline{u}_{\mu;\alpha}\:\Lambda^{\alpha}$ or $\Lambda_{\mu}\:\Lambda_{\nu;\alpha}\:\Lambda^{\alpha}+\Lambda_{\nu}\:\Lambda_{\mu;\alpha}\:\Lambda^{\alpha}$ just to show a few examples. The tensor $\overline{\tau}^{\mu\nu}(\overline{u})$ is just (\ref{SETIMPVISC}) and we can write $\overline{\tau}^{\mu\nu}(\overline{u}+\Lambda)=\overline{\tau}^{\mu\nu}(\overline{u})+\overline{\tau}^{\mu\nu}(\Lambda)$. Next we apply in expression (\ref{SETIMP}) the simultaneous transformations $\overline{u}^{\alpha} \rightarrow \overline{u}^{\alpha}+\Lambda^{\alpha}$ plus transformations (\ref{rhotransf}-\ref{viscoustransf}). We find,

\begin{eqnarray}
\overline{T}^{imp\:\mu\nu} \rightarrow \overline{T}^{imp\:\mu\nu} &+& (\overline{\rho} + \overline{p})\:(\overline{u}^{\mu}\:\Lambda^{\nu} + \overline{u}^{\nu}\:\Lambda^{\mu} + \Lambda^{\mu}\:\Lambda^{\nu}) \nonumber \\ &+& (\widetilde{\rho} + \widetilde{p})\:(\overline{u}^{\mu}+\Lambda^{\mu})\:(\overline{u}^{\nu}+\Lambda^{\nu}) + \widetilde{p}\:\overline{g}^{\mu\nu}
+ \Lambda^{\mu}\:\overline{q}^{\nu} + \Lambda^{\nu}\:\overline{q}^{\mu} + \nonumber \\ &+& \widetilde{q}^{\mu}\:(\overline{u}^{\nu}+\Lambda^{\nu}) + \widetilde{q}^{\nu}\:(\overline{u}^{\mu}+\Lambda^{\mu}) + \overline{\tau}^{\mu\nu}(\Lambda) + \widetilde{\tau}^{\mu\nu} \ . \label{eqTimp}
\end{eqnarray}

The next step is imposing invariance,

\begin{eqnarray}
&&(\overline{\rho} + \overline{p})\:(\overline{u}^{\mu}\:\Lambda^{\nu} + \overline{u}^{\nu}\:\Lambda^{\mu} + \Lambda^{\mu}\:\Lambda^{\nu}) \nonumber \\ &+& (\widetilde{\rho} + \widetilde{p})\:(\overline{u}^{\mu}+\Lambda^{\mu})\:(\overline{u}^{\nu}+\Lambda^{\nu}) + \widetilde{p}\:\overline{g}^{\mu\nu} + \Lambda^{\mu}\:\overline{q}^{\nu} + \Lambda^{\nu}\:\overline{q}^{\mu} \nonumber \\ &+& \widetilde{q}^{\mu}\:(\overline{u}^{\nu}+\Lambda^{\nu}) + \widetilde{q}^{\nu}\:(\overline{u}^{\mu}+\Lambda^{\mu}) + \overline{\tau}^{\mu\nu}(\Lambda) + \widetilde{\tau}^{\mu\nu} = 0 \ . \label{eqTimpinv}
\end{eqnarray}

We contract equation (\ref{eqTimpinv}) with $(\overline{u}_{\nu}+\Lambda_{\nu})$ and use equations (\ref{conditionsheat}-\ref{conditionsvisc}) in order to find the object $\widetilde{q}^{\mu}$,
\begin{eqnarray}
\widetilde{q}^{\mu} &=& \{(\overline{q}^{\alpha}\:\Lambda_{\alpha})\:(\overline{u}^{\mu}+\Lambda^{\mu}) + \overline{\tau}^{\mu\alpha}(\overline{u})\:\Lambda_{\alpha} \nonumber \\ &-& [(\overline{\rho} + \overline{p})\:(\overline{u}^{\mu}\:\Lambda^{\alpha} + \overline{u}^{\alpha}\:\Lambda^{\mu} + \Lambda^{\mu}\:\Lambda^{\alpha}) + \Lambda^{\mu}\:\overline{q}^{\alpha} + \Lambda^{\alpha}\:\overline{q}^{\mu} \nonumber \\ &+& (\widetilde{\rho} + \widetilde{p})\:(\overline{u}^{\mu}+\Lambda^{\mu})\:(\overline{u}^{\alpha}+\Lambda^{\alpha}) + \widetilde{p}\:\overline{g}^{\mu\alpha}]\:(\overline{u}_{\alpha}+\Lambda_{\alpha})\}/
(\overline{u}^{\nu}+\Lambda^{\nu})\:(\overline{u}_{\nu}+\Lambda_{\nu}) \ . \label{qtilde}
\end{eqnarray}

Contracting again with $(\overline{u}_{\mu}+\Lambda_{\mu})$ and using one more time the condition (\ref{conditionsheat}) we can obtain after some straightforward algebraic work the equation corresponding to $\widetilde{\rho}$ and $\widetilde{p}$,



\begin{eqnarray}
&&(\widetilde{\rho} + \widetilde{p})\:[(\overline{u}^{\mu}+\Lambda^{\mu})\:(\overline{u}_{\mu}+\Lambda_{\mu})]^{2}
+ \widetilde{p}\:\overline{g}^{\mu\alpha}\:(\overline{u}_{\alpha}+\Lambda_{\alpha})\:(\overline{u}_{\mu}+\Lambda_{\mu}) = \nonumber \\ &&\{(\overline{q}^{\alpha}\:\Lambda_{\alpha})\:(\overline{u}^{\mu}+\Lambda^{\mu})\:(\overline{u}_{\mu}+\Lambda_{\mu}) + \overline{\tau}^{\mu\alpha}(\overline{u})\:\Lambda_{\alpha}\:(\overline{u}_{\mu}+\Lambda_{\mu}) \nonumber \\ &-& [(\overline{\rho} + \overline{p})\:(\overline{u}^{\mu}\:\Lambda^{\alpha} + \overline{u}^{\alpha}\:\Lambda^{\mu} + \Lambda^{\mu}\:\Lambda^{\alpha}) + \Lambda^{\mu}\:\overline{q}^{\alpha} + \Lambda^{\alpha}\:\overline{q}^{\mu}]\: . \nonumber \\ &.& (\overline{u}_{\alpha}+\Lambda_{\alpha})\:(\overline{u}_{\mu}+\Lambda_{\mu})\} - (\overline{u}^{\nu}+\Lambda^{\nu})\:(\overline{u}_{\nu}+\Lambda_{\nu})\:(-\overline{q}^{\alpha}\:\Lambda_{\alpha}) \ . \label{ptilde3}
\end{eqnarray}

We know this equation of state $\widetilde{p}(\widetilde{\rho})$ from the outset independently from the above analysis, and we also get to find $\widetilde{\rho}$ just by equating these two independent expressions of $\widetilde{p}(\widetilde{\rho})$ (the one we know from the outset plus the one from equation (\ref{ptilde3})) and then replacing both $\widetilde{p}(\widetilde{\rho})$ and $\widetilde{\rho}$ back in equation (\ref{qtilde}). Once $\widetilde{q}^{\mu}$ is found in equation (\ref{qtilde}) we replace it along with $\widetilde{p}(\widetilde{\rho})$ and $\widetilde{\rho}$ in equation (\ref{eqTimpinv}) and find $\widetilde{\tau}^{\mu\nu}$.
We start with the perturbed imperfect stress-energy tensor (\ref{SETIMP}) and then we implement simultaneous transformations $\overline{u}^{\alpha} \rightarrow \overline{u}^{\alpha}+\Lambda^{\alpha}$ plus corresponding transformations (\ref{rhotransf}-\ref{viscoustransf}). The original (\ref{SETIMP}) plus the four-velocity gauge-like transformation scalar $\Lambda$ or the gradient $\Lambda_{\mu}$ are all known from the outset as are also known from the start the equations of state $\overline{p}(\overline{\rho})$, $\widetilde{p}(\widetilde{\rho})$ and $\widetilde{p}(\overline{p}(\overline{\rho}),\overline{\rho},\widetilde{\rho})$. In fact from equation (\ref{ptilde3}) we observe that $\widetilde{p}$ is $\widetilde{p}(\overline{p}(\overline{\rho}),\overline{\rho},\widetilde{\rho})$. These are the two expressions for $\widetilde{p}$ that we equate in order to obtain $\widetilde{\rho}$ in terms of the objects given at the outset. Next we impose invariance through equations (\ref{eqTimpinv}) under all of these four-velocity gauge-like local transformations. Finally and using the conditions (\ref{conditionsheat}-\ref{conditionsvisc}) we obtain by algebraic work both $\widetilde{q}^{\mu}$ and $\widetilde{\tau}^{\mu\nu}$ plus $\widetilde{p}(\widetilde{\rho})$ and $\widetilde{\rho}$. We have available ten equations or invariance conditions through equations (\ref{eqTimpinv}) plus the five conditions (\ref{conditionsheat}-\ref{conditionsvisc}) and found fifteen local objects $\widetilde{q}^{\mu}$, $\widetilde{\tau}^{\mu\nu}$ and $\widetilde{p}(\widetilde{\rho})$ or $\widetilde{\rho}$. This system of ideas ensures the gauge-like invariance in the right hand side of the system (\ref{EFEIMP}) knowing from the outset that the left hand side of the system (\ref{EFEIMP}) is explicitly and manifestly invariant under this local group of transformations since this latter claim has been proven in sections I-II-III of reference \cite{AGNS} and separately also in reference \cite{A}. It is also clear that the conservation equations $\overline{\nabla}_{\nu}\overline{T}^{imp\:\mu\nu}=0$ are also invariant under the new local symmetry because on one hand the stress-energy tensor $\overline{T}^{imp\:\mu\nu}$ is invariant and on the other hand the perturbed metric tensor $\overline{g}_{\mu\nu}$ is invariant under this local symmetry and therefore the Christoffel symbols are also invariant.

\section{Comments on the stress-energy perturbed vorticity tensor}
\label{vorticity}

In section \ref{invariance} a thorough discussion is provided on how a perturbed imperfect fluid stress-energy tensor can be invariant under gauge-like four-velocity local transformations $\overline{u}^{\alpha} \rightarrow \overline{u}^{\alpha}+\Lambda^{\alpha}$. It is concluded that specific transformations have to be simultaneously satisfied for the perturbed heat flow currents, the viscous stresses, the density and pressure for the right hand side of the Einstein equations for the perturbed imperfect fluid to be invariant under this kind of local transformation.
We would also like to proceed in complete analogy to the electromagnetic case presented in reference \cite{A} and also reference \cite{AGNS} for the unperturbed case in order to present a proposal for this perturbed vorticity stress-energy tensor and to the study of its properties, see references \cite{GKB,TQH,CD,GQW,MSS}. Let us remember that the unperturbed vorticity stress-energy tensor was presented in paper \cite{AGNS}. Let us proceed to introduce the following symmetric tensor,

\begin{equation}
\overline{T}^{vort}_{\mu\nu}=\overline{\xi}_{\mu\lambda}\:\:\overline{\xi}_{\nu}^{\:\:\:\lambda}
+ \ast \overline{\xi}_{\mu\lambda}\:\ast \overline{\xi}_{\nu}^{\:\:\:\lambda}\ .\label{TEMDR}
\end{equation}

When we consider equations (\ref{i2}) and (\ref{scond}) it becomes simple to prove that the tetrad sets (\ref{V1}-\ref{V4}) and (\ref{Uw}-\ref{Ww}) diagonalize locally and covariantly the stress-energy tensor (\ref{TEMDR}). We consider this tensor to be a stress-energy tensor because it is built in the same way as in Einstein-Maxwell spacetimes just replacing the electromagnetic four-potential by the four-velocity. Using the inverse of the local duality rotation introduced in equation (\ref{vef}) and given by $\overline{u}_{[\mu;\nu]} = \cos\overline{\alpha} \:\: \overline{\xi}_{\mu\nu} + \sin\overline{\alpha} \:\: \ast \overline{\xi}_{\mu\nu}$ we get,

\begin{equation}
\overline{T}^{vort}_{\mu\nu}=\overline{u}_{[\mu;\lambda]}\:\:\overline{u}_{[\nu;\rho]}\:\overline{g}^{\rho\lambda}
+ \ast \overline{u}_{[\mu;\lambda]}\:\ast \overline{u}_{[\nu;\rho]}\:\overline{g}^{\rho\lambda}=\overline{\xi}_{\mu\lambda}\:\:\overline{\xi}_{\nu}^{\:\:\:\lambda}
+ \ast \overline{\xi}_{\mu\lambda}\:\ast \overline{\xi}_{\nu}^{\:\:\:\lambda} \ .\label{TEMDRINV}
\end{equation}

Leaving for the moment possible constant units factors aside it is also clear that this tensor is not the whole stress-energy tensor on the right hand side of the perturbed Einstein fluid equations. The complete tensor fulfilling the conservation equations $\overline{\nabla}_{\nu}\overline{T}^{\mu\nu}=0$ would include the perturbed perfect fluid terms plus perturbed heat flow plus perturbed viscous stress plus the perturbed vorticity stress-energy, see section \ref{conclusions}. Vectors (\ref{V1}-\ref{V2}) or the normalized (\ref{Uw}-\ref{Vw}) generate the local plane one where all vectors are eigenvectors of the tensor (\ref{TEMDR}) with eigenvalue $\overline{Q}/2=\overline{\xi}_{\mu\lambda}\:\overline{\xi}_{\mu}^{\:\:\:\lambda}/2$ which we assume to be $\overline{Q}\neq0$. Vectors (\ref{V3}-\ref{V4}) or the normalized (\ref{Zw}-\ref{Ww}) generate the local orthogonal plane two where all vectors are eigenvectors of the tensor (\ref{TEMDR}) with eigenvalue $-\overline{Q}/2=-\overline{\xi}_{\mu\lambda}\:\overline{\xi}_{\mu}^{\:\:\:\lambda}/2$. Under the transformation $\overline{u}^{\alpha} \rightarrow \overline{u}^{\alpha}+\Lambda^{\alpha}$ the vectors that span the local plane one would undergo LB1 transformations inside this plane while the vectors that span the local plane two would undergo $SO(2)$ transformations inside this second plane. Since the perturbed curl field $\overline{u}_{[\mu;\nu]}$ is locally invariant under this transformation, the perturbed vorticity stress-energy (\ref{TEMDRINV}) is invariant as well. The electromagnetic case analyzed in reference \cite{A} is identical in mathematical structure to our present vorticity case. The principle of symmetry is the hidden principle that has been guiding us in our search for the vorticity stress-energy symmetric tensor.

\section{Conclusions}
\label{conclusions}

In a previous manuscript \cite{AGNS} it was found a new symmetry for an imperfect fluid in curved four-dimensional Lorentz spacetimes. The key element in this discovery was that when there is vorticity we can use the curl of the four-velocity as an antisymmetric field in order to build new tetrads with remarkable properties. Through a local duality transformation by a local angle that we called the complexion in equation (\ref{vef}) it was found an extremal field that enabled the construction of tetrad skeletons. These new tetrads that diagonalize the  perfect fluid stress-energy tensor and the vorticity stress-energy tensor \cite{AGNS} as well have two construction components, the skeletons and the gauge vectors. The gauge vectors are a choice that we can make and we chose the gauge vectors to be the four-velocity. There arose the idea of four-velocity gauge-like transformations, because we can also gauge the tetrads with the four-velocity plus a gradient and see how the tetrads transform. These new tetrads define at every point in spacetime two orthogonal planes. The timelike-spacelike plane or plane one and the spacelike-spacelike plane or plane two. The tetrad vectors that span these local planes transform under four-velocity gauge-like transformations in such a way that they do not leave these planes after the transformation. In fact it was previously proved in manuscript \cite{A} that the local group of Abelian four-velocity gauge-like transformations in plane one is isomorphic to the group LB1 of tetrad transformations and in plane two to the group LB2 of tetrad transformations. The group LB1 is given by $SO(1,1) \times Z_{2} \times Z_{2}$ where $SO(1,1)$ is proper orthochronous. The first $Z_{2}$ is given by $\{I_{2 \times 2}, -I_{2 \times 2}\}$ and the second $Z_{2}$ is given by $\{I_{2 \times 2}, \mbox{the swap}\: (01|10)\}$. We would have to add in order to complete the image of the map $SO(1,1) \times Z_{2} \times Z_{2}\: \bigoplus \: \{light\:cone\:gauge\}$ where the light cone gauge includes the four solutions to the differential equations in the local future and past light cones established in manuscripts \cite{A,AGE,ROMP,SING}. One of these discrete transformations is the full inversion or minus the identity two by two. It is a Lorentz transformation and we designated this discrete transformation above by the notation, $-I_{2 \times 2}$. The other discrete transformation is given by $\Lambda^{o}_{\:\:o} = 0$, $\Lambda^{o}_{\:\:1} = 1$, $\Lambda^{1}_{\:\:o} = 1$,  $\Lambda^{1}_{\:\:1} = 0$, which is not a Lorentz transformation because it is a reflection, see reference \cite{A} for the whole analysis. We designated this discrete transformation above by the notation, $\mbox{the swap}\: (01|10)$. The local group of Abelian four-velocity gauge-like transformations is proven independently to be isomorphic to the local group LB2 of tetrad transformations on the local plane two and it is $SO(2)$. We can mention applications in relativistic astrophysics such as \cite{SSAK}:

\begin{enumerate}

\item Regarding perturbations to imperfect fluids in cosmology we can cite reference \cite{SSAK} and we quote ``We present a new prescription for analysing cosmological perturbations in a more general class of scalar-field dark-energy models where the energy-momentum tensor has an imperfect-fluid form. This class includes Brans-Dicke models, $f(R)$ gravity, theories with kinetic gravity braiding and generalised galileons. We employ the intuitive language of fluids, allowing us to explicitly maintain a dependence on physical and potentially measurable properties. We demonstrate that hydrodynamics is not always a valid description for describing cosmological perturbations in general scalar-field theories and present a consistent alternative that nonetheless utilises the fluid language. We apply this approach explicitly to a worked example: k-essence non-minimally coupled to gravity. This is the simplest case which captures the essential new features of these imperfect-fluid models. We demonstrate the generic existence of a new scale separating regimes where the fluid is perfect and imperfect. We obtain the equations for the evolution of dark-energy density perturbations in both these regimes. The model also features two other known scales: the Compton scale related to the breaking of shift symmetry and the Jeans scale which we show is determined by the speed of propagation of small scalar-field perturbations, i.e. causality, as opposed to the frequently used definition of the ratio of the pressure and energy-density perturbations.''

\item Let us see for example a simple result from neutron star perfect fluid approximations. We quote from reference \cite{SCNO} ``The macroscopic neutron vorticity $\varpi_{n}$ (we use Greek letters for spacetime indices) $\varpi_{n}=\sqrt{\frac{\varpi_{\mu\nu}\:\varpi^{\mu\nu}}{2}}$ where the vorticity 2-form $\varpi_{n}$ is defined by $\varpi_{\mu\nu}=\nabla_{\mu}p_{\nu}^{n}-\nabla_{\nu}p_{\mu}^{n}$, $p_{\mu}^{n}$ denoting the conjugate superfluid momentum. We note here that, on length scales smaller than the intervortex separation $d_{v}$, typically of the order of $d_{v} \sim n_{v}^{-1/2} \sim 10^{-3} cm$ (see Eq. (1)), $\varpi_{n}$ strictly vanishes because $p_{\mu}^{n}$ should be locally proportional to the gradient of a quantum scalar phase. Nevertheless, on the large scales we are interested in here, the neutron vorticity 2-form is non-vanishing, as well as its corresponding scalar amplitude $\varpi_{n}$'', see references \cite{SCNO,CH,LP,HM,BC1,LSC,CPG,EGU,ASC,SON,CS,CC,PA}. Since a neutron star might accrete matter from a companion star for many years the neutron-star crust might be set out of its thermal equilibrium with the core. After the accretion ends the heated crust relaxes towards a state of equilibrium. The time of thermal relaxation depends specially on the crust heat capacity. Were the neutrons not superfluid they would be able to store so much heat that the thermal relaxation would last longer than the observed times. However, the thermal relaxation mechanism of these systems is not completely understood. For instance, additional heat sources unknown so far need to be considered in order to reproduce accurately the observations \cite{DWB,TAP}. We wonder if this discrepancy might be related to the non-consideration of a vorticity stress-energy tensor. In what follows we will consider the components of the Ricci tensor when we can neglect the imperfect fluid terms in the perturbed equation (\ref{SETIMP}) and also in equation (\ref{SETIMPVISC}). We will only keep the perturbed perfect fluid terms plus the perturbed vorticity terms. We are not saying that the local four-velocity gauge-like symmetry that we have been studying in this manuscript is no longer valid, rather we consider that even though the symmetry still exists, for practical purposes the perturbed stress-energy tensor imperfect fluid terms can be dismissed or neglected with respect to the perturbed perfect fluid stress-energy terms for example. In neutron star physics there is ample literature, we just cite some papers and reviews where more references can be found therein \cite{F8,NS,BC,RS,NC,AC2,FI,AWV,DL,291}. We apply this new technique using only local algebraic covariant analysis which will not add any more substantial computational time and however simplify further applications like in spacetime dynamical evolution. The stress-energy tensor under all of the above simplifications can be expressed by,

\begin{equation}
\overline{T}_{\mu\nu} = (\overline{\rho} + \overline{p})\:\overline{u}_{\mu}\:\overline{u}_{\nu} + \overline{p}\:\overline{g}_{\mu\nu} + \overline{T}^{vort}_{\mu\nu}\ ,\label{SETVORT}
\end{equation}

     where the vorticity stress-energy tensor is presented in equation (\ref{TEMDR}). If we call $\overline{Q}^{vort}=\overline{\xi}_{\mu\lambda}\:\:\overline{\xi}^{\mu\lambda}$ and we keep in mind that the metric tensor is given by equation (\ref{MT}) we can write the tensor (\ref{SETVORT}) as,

\begin{equation}
\overline{T}_{\mu\nu} = (\overline{\rho} + \overline{p})\:\overline{u}_{\mu}\:\overline{u}_{\nu} + \overline{p}\:\overline{g}_{\mu\nu} + \frac{\overline{Q}^{vort}}{2}\:[-\hat{\overline{U}}_{\mu}\:\hat{\overline{U}}_{\nu} + \hat{\overline{V}}_{\mu}\:\hat{\overline{V}}_{\nu} -
\hat{\overline{Z}}_{\mu}\:\hat{\overline{Z}}_{\nu} - \hat{\overline{W}}_{\mu}\:\hat{\overline{W}}_{\nu}] \ .\label{SETVORTEXP}
\end{equation}

      We can also check using the identity equations (\ref{i2}) and (\ref{scond}) that,

\begin{eqnarray}
\hat{\overline{U}}^{\alpha}\:\overline{T}_{\alpha}^{\:\:\:\beta} &=& (\overline{\rho} + \overline{p})\:(\hat{\overline{U}}^{\alpha}\:\overline{u}_{\alpha})\:\overline{u}^{\beta} + [\overline{p}+\frac{\overline{Q}^{vort}}{2}]\:\hat{\overline{U}}^{\beta}
\label{PVSETU}\\
\hat{\overline{V}}^{\alpha}\:\overline{T}_{\alpha}^{\:\:\:\beta} &=& [\overline{p}+\frac{\overline{Q}^{vort}}{2}]\:\hat{\overline{V}}^{\beta}
\label{PVSETV}\\
\hat{\overline{Z}}^{\alpha}\:\overline{T}_{\alpha}^{\:\:\:\beta} &=& [\overline{p}+\frac{\overline{Q}^{vort}}{2}]\:\hat{Z}^{\beta}
\label{PVSETZ}\\
\hat{\overline{W}}^{\alpha}\:\overline{T}_{\alpha}^{\:\:\:\beta} &=&  (\overline{\rho} + \overline{p})\:(\hat{\overline{W}}^{\alpha}\:\overline{u}_{\alpha})\:\overline{u}^{\beta} + [\overline{p}-\frac{\overline{Q}^{vort}}{2}]\:\hat{\overline{W}}^{\beta}\ .
\label{PVSETW}
\end{eqnarray}

     We can also introduce the expression $\overline{u}^{\beta}=-(\hat{\overline{U}}^{\alpha}\:\overline{u}_{\alpha})\:\hat{\overline{U}}^{\beta}+
     (\hat{\overline{W}}^{\alpha}\:\overline{u}_{\alpha})\:\hat{\overline{W}}^{\beta}$ and from equations (\ref{Vw}-\ref{Zw}) we know that $\hat{\overline{V}}^{\alpha}\:\overline{u}_{\alpha}=0$ and $\hat{\overline{Z}}^{\alpha}\:\overline{u}_{\alpha}=0$. Using these results and the equations (\ref{PVSETU}-\ref{PVSETW}) we can immediately conclude that only five components of the stress-energy tensor (\ref{SETVORT}) or the equivalent (\ref{SETVORTEXP}) are not zero and this is a relevant set of simplifications. The tetrad that we have found (\ref{Uw}-\ref{Ww}) will be used in order to find if there is a possible simplification on the left hand side of the Einstein equations independently of our stress-energy tensor procedure of diagonalization on the right hand side of the Einstein equations for perturbed fields. We consider the following equation that can be found in the bibliography \cite{CW,WE},

\begin{eqnarray}
v_{\mu;\nu;\rho}-v_{\mu;\rho;\nu} = -\overline{R}^{\sigma}_{\:\:\:\mu\nu\rho}\:v_{\sigma} \ . \label{dcdanti}
\end{eqnarray}

    Equation (\ref{dcdanti}) is valid in general for any vector field $v_{\sigma}$. In order to verify a particular example of simplification let us use the previous algorithm starting with equation (\ref{dcdanti}) applied to the vector (\ref{Ww}) after contracting this equation on both sides with $\overline{g}^{\mu\rho}$,

\begin{eqnarray}
\hat{\overline{W}}^{\mu}_{\:\:\:;\nu;\mu}-\hat{\overline{W}}^{\mu}_{\:\:\:;\mu;\nu} = -\overline{R}^{\sigma}_{\:\:\:\nu}\:\hat{\overline{W}}_{\sigma} \ . \label{dcdantikillVR}
\end{eqnarray}

    Finally, we contract with vector (\ref{Zw}),

\begin{eqnarray}
\hat{\overline{Z}}^{\nu}\:\hat{\overline{W}}^{\mu}_{\:\:\:;\nu;\mu}-\hat{\overline{Z}}^{\nu}\:\hat{\overline{W}}^{\mu}_{\:\:\:;\mu;\nu} = -\hat{\overline{Z}}^{\nu}\:\overline{R}^{\sigma}_{\:\:\:\nu}\:\hat{\overline{W}}_{\sigma} = -\hat{\overline{Z}}^{\nu}\:(\overline{T}_{\:\:\:\nu}^{\sigma}-
\frac{1}{2}\:\delta^{\sigma}_{\:\:\:\nu}\:\overline{T}^{\mu}_{\:\:\:\mu})\:\hat{\overline{W}}_{\sigma} = 0 \ , \label{EEVZ}
\end{eqnarray}

    since $\hat{\overline{Z}}^{\alpha}\:\overline{T}_{\alpha}^{\:\:\:\beta} = [\overline{p}+\frac{\overline{Q}^{vort}}{2}]\:\hat{\overline{Z}}^{\beta}$ according to equation (\ref{PVSETZ}) and also using the orthogonality $\hat{\overline{Z}}^{\sigma}\:\hat{\overline{W}}_{\sigma} = 0$. Equation (\ref{EEVZ}) is another source of simplification. This idea can be repeated with other components as well for this perturbed formulation. We notice that all the simplifications brought about by the new tetrads in the perturbed case as compared to the unperturbed case, remain. Resuming our original analysis, it was proved that the local group of Abelian four-velocity gauge-like transformations was a symmetry of the metric tensor. Therefore, a symmetry of the Einstein equations left hand side. It was concluded that it should also be a symmetry of the Einstein equations right hand side and a symmetry of the imperfect fluid stress-energy tensor. It is not evidently a symmetry of the perfect fluid stress-energy tensor but it was found to be a symmetry of the imperfect fluid stress-energy tensor. The perfect fluid stress-energy tensor will not be invariant under gauge-like four-velocity transformations by itself. In order to make it invariant we would have to add to the right hand side of the Einstein equations in (\ref{EFE}) terms including heat flow currents $(\overline{\rho} + \overline{p})\:(\overline{u}_{\mu}\:\overline{q}_{\nu} + \overline{u}_{\nu}\:\overline{q}_{\mu})$ where $\overline{u}^{\mu}\:\overline{q}_{\mu}=0$ is satisfied and terms with viscous stresses in the fluid $\overline{\tau}_{\mu\nu}$ where $\overline{u}^{\mu}\:\overline{\tau}_{\mu\nu}=0$ is also satisfied, see references \cite{HS,CW,RW} and section \ref{invariance}. The whole symmetry proof of the imperfect fluid was presented in manuscript \cite{AGNS}. In this present manuscript we have developed the proof for a source and gravitational field of imperfect fluid under perturbations.

\end{enumerate}

Once again we must stress that even if the notation and the paper structure are similar to reference \cite{AGNS} the content is different because unlike manuscript \cite{AGNS} in this manuscript we are studying imperfect fluids under perturbations. We have proven the dynamic nature of local symmetries and this proof requires a similar paper structure and notation with respect to reference \cite{AGNS} but the content is substantially different. The instantaneous symmetry is the object of our study. We proved that the local orthogonal planes one and two tilt under perturbations continuous or discrete and that the symmetries become instantaneous. The whole system of ideas for the unperturbed imperfect fluids, see reference \cite{AGNS},  was proved in this manuscript to be reproduced for the perturbed scenario. The key to understand the concept that we are presenting in this manuscript is that there is for fluids with vorticity a sector of analogous ideas as presented in the paper \cite{A} for the electromagnetic and gravitational fields in Einstein-Maxwell spacetimes. In Einstein-Maxwell spacetimes there is a non-trivial curl of the electromagnetic potential four-vector. There is manifest electromagnetic gauge invariance of the metric tensor as expressed in terms of tetrads of an analogous nature as to (\ref{V1}-\ref{V4}) or (\ref{Uw}-\ref{Ww}). The main difference between the Einstein-Maxwell spacetimes and the perfect fluid spacetimes is the stress-energy tensor. In Einstein-Maxwell spacetimes the stress-energy tensor is invariant under electromagnetic gauge transformations while in perfect fluid spacetimes the stress-energy tensor is not necessarily invariant under local gauge-like transformations of the four-velocity vectors. It is necessary to introduce heat flows and viscous stresses to make it invariant. We proceeded in this direction in manuscript \cite{AGNS} where we introduced a new kind of local transformation in the heat flux vector, the viscous stress-energy tensor, the density and pressure in order to make the whole imperfect fluid stress-energy tensor invariant. In this present manuscript we proved that for perturbed formulations the basic constructions, properties and symmetries, remain instantaneously, see also \cite{LPAP,MON,GP,dsmg,dsmg2}. The summary of all these findings can be stated in the following theorem,

\newtheorem {guesslb1} {Theorem}
\begin{guesslb1}
The local four-velocity gauge-like symmetry already found for imperfect fluids with vorticity becomes instantaneous under perturbations. The local planes of symmetry tilt under perturbations and even though the symmetries are continuously or discretely broken at the points in spacetime, new symmetries arise. There is a local symmetry evolution. \end{guesslb1}

\section{Declaration of interest statement}
\label{interest}

The authors declare that they have no known competing financial interests or personal relationships that could have appeared to influence the work reported in this paper.

\section{Data availability statement}
\label{data}

There is no data to be reported in this paper.


\begin{thebibliography}{spmpsci}

\bibitem{AGNS} A. Garat, ``New symmetry for the imperfect fluid'', Eur. Phys. J. C, {\bf 80} 4 (2020) 333. https://doi.org/10.1140/epjc/s10052-020-7887-9

\bibitem{AGNSHC} A. Garat, New symmetry in higher curvature spacetimes, (2022) EPL 137 19003.
https://iopscience.iop.org/article/10.1209/0295-5075/ac525c/meta

\bibitem{EG} E. Gourgoulhon, {\it Proceedings of the School Astrophysical Fluid Dynamics, Carg\`{e}se, France}
(EDP Sciences, 2006).

\bibitem{AC} N. Andersson and G. L. Comer, Living Rev. Relativity, {\it Relativistic Fluid Dynamics: Physics for many different scales} (2007). http://www.livingreviews.org/lrr-2007-1.

\bibitem{F} J. A. Font, Living Rev. Relativity, {\it Numerical Hydrodynamics in General Relativity} (2003). http://www.livingreviews.org/lrr-2003-4.

\bibitem{JO} J. Olshoorn, {\it Relativistic fluid dynamics}, The Waterloo Mathematics Review, Vol. {\bf 1} Issue {\bf 2} (2011).

\bibitem{A} A. Garat, Tetrads in geometrodynamics, J. Math. Phys. {\bf 46}, 102502 (2005).
\bibitem{AGE} A. Garat, Erratum: Tetrads in geometrodynamics, J. Math. Phys. {\bf 55}, 019902 (2014).

\bibitem{ROMP} A. Garat, Isomorphism Between the Local Poincar\'{e} Generalized Translations Group and the Group of Spacetime Transformations $(\bigotimes LB1)^{4}$, Reports on Mathematical Physics, Volume {\bf 86}, Issue 3, December 2020, Pages 355-382.

\bibitem{SING} A. Garat, ``Singular gauge transformations in geometrodynamics'', Int. J. Geom. Methods Mod. Phys., Vol. {\bf 18}, No. 10, (2021) 2150150 (35 pages) World Scientific Publishing Company. DOI: 10.1142/S0219887821501504.

\bibitem{ATGU} A. Garat, Einstein-Maxwell tetrad grand unification, Int. J. Geom. Methods Mod. Phys., Vol. {\bf 17} , No. {\bf 8} (2020) 2050125. DOI: S021988782050125X.

\bibitem{IWCP} A. Garat, {\it New tetrads in Riemannian geometry and new ensuing results in group theory, gauge theory and fundamental physics in particle physics, general relativity and astrophysics}, Int. J. Mod. Phys. Conf. Ser., Vol. {\bf 45}, (2017), 1760004.

\bibitem{LomCon} A. Garat, {\it Local Groups of Internal Transformations Isomorphic to Local Groups of Spacetime Tetrad Transformations}, World Scientific, Particle Physics at the Silver Jubilee of Lomonosov Conferences, pp. 510-514 (2019).

\bibitem{AEO} A. Garat, Euler observers in geometrodynamics, Int. J. Geom. Meth. Mod. Phys., Vol. {\bf 11} (2014), 1450060. arXiv:gr-qc/1306.4005

\bibitem{ACD} A. Garat, Covariant diagonalization of the perfect fluid stress-energy tensor, Int. J. Geom. Meth. Mod. Phys., Vol. {\bf 12} (2015), 1550031. arXiv:gr-qc/1211.2779

\bibitem{ENV} A. Garat, Euler observers for the perfect fluid without vorticity, Z. Angew. Math. Phys. (2019) 70: 119.

\bibitem{MW} C. Misner and J. A. Wheeler, Classical physics as geometry, Annals of Physics {\bf 2}, 525 (1957).

\bibitem{SSAK} I. Sawickia, I. D. Saltasb, L. Amendolaa and M. Kunzd, {\it Consistent perturbations in an imperfect fluid}, Journal of Cosmology and Astroparticle Physics, Volume 2013, January 2013.

\bibitem{EC} E. Calzetta, {\it Relativistic fluctuating hydrodynamics}, Class. Quantum Grav. {\bf 15} Number {\bf 3}, 653 (1998).

\bibitem{LMRB} J. A. López Molina, M. J. Rivera and E. Berjano, {\it Fourier, hyperbolic and relativistic heat transfer equations: a comparative analytical study}, Proceedings of the Royal Society A, Volume 470, Issue 2172 (2014).

\bibitem{HS} H.~Stephani, {\it General Relativity} (Cambridge University Press, Cambridge, 2000).

\bibitem{CW} I. Ciufolini and J. A. Wheeler, {\it Gravitation and Inertia} (Princeton University Press, 1995).

\bibitem{RW} R. ~Wald, {\it General Relativity} (University of Chicago Press, Chicago, 1984).

\bibitem{GKB} G. K. Batchelor,  {\it An introduction to fluid dynamics} (Cambridge University Press, Cambridge, 2009).

\bibitem{TQH} R. E. Wyatt, {\it Topics in quantum hydrodynamics: The stress-energy tensor and vorticity}, Quantum Dynamics with Trajectories pp 322-353. https://doi.org/10.1007/0-387-28145-2-13. (Part of the interdisciplinary applied mathematics book series, Springer, New York, 2009).

\bibitem{CD} M. Czubak and M. M. Disconzi, ``On the well posedness of relativistic viscous fluids with non-zero vorticity'', J. Math. Phys. {\bf 57}, 042501 (2016). arXiv:1407.6963[math-ph].

\bibitem{GQW} J. H. Gao, B. Qi and S. Y. Wang, ``Vorticity and magnetic field production in relativistic ideal fluids'', Phys. Rev. {\bf D90} no. 8, 083001 (2015). arXiv:1406.1944[hep-ph].

\bibitem{MSS} M. S. Swanson, {\it Classical field theory and the stress-energy tensor}, (A Morgan and Claypool Publication as part of IOP Concise Physics, San Rafael, CA, USA, 2015).

\bibitem{SCNO} A. Sourie, N. Chamel, J. Novak and M. Oertel, Global numerical simulations of vortex-mediated pulsar glitches in full general relativity, Mon. Not. R. Astr. Soc., {\bf 464} Issue 4 pp 4641-57 (2016). ({\it Preprint} arXiv:1607.08213 [astro.ph-HE])

\bibitem{CH} N. Chamel and P. Haensel, Entrainment parameters in cold superfluid neutron star core, Phys. Rev. C, {\bf 73} 045802 (2006). ({\it Preprint} arXiv:nucl-th/0603018)

\bibitem{LP} L. Lehner and F. Pretorius, {\it Numerical relativity and astrophysics}, Annual Review of Astronomy and Astrophysics, {\bf 52} pp 661-94 (2017). ({\it Preprint} arXiv:1405.4840 [astro.ph-HE])

\bibitem{HM} B. Haskell and A. Melatos, Models of pulsar glitches, Int. J. Mod. Phys. D {\bf 24} No. 03 1530008 (2015). ({\it Preprint} arXiv:1502.07062 [astro.ph-SR])

\bibitem{BC1} B. Carter, Relativistic superfluid models for rotating neutron stars (ECT workshop, Trento, June 2000) ({\it Preprint} arXiv:astro-ph/0101257)

\bibitem{LSC} D. Langlois, D. M. Sedrakian and B. Carter, Differential rotation of relativistic superfluid in neutron stars, Mon. Not. R. Astr. Soc., {\bf 297} Issue 4 pp 1189-1201 (1998). ({\it Preprint} arXiv:astro-ph/9711042)

\bibitem{CPG} N. Chamel, J. M. Pearson and S. Goriely, Superfluidity and entrainment in neutron-star crust {\it Publications of the Astronomical Society of the Pacific}  {\bf 466:203} ({\it Proc.} of the ERPM {\it Conf.}, Zielona Gora, Poland) (2012) ({\it Preprint} arXiv:1206.6926 [astro-ph.HE])

\bibitem{EGU} E. G\"{u}gercino\u{g}lu, Vortex creep against toroidal flux lines, crustal entrainment and pulsar glitches, ApJ,  {\bf 788} L11 (2014). ({\it Preprint} arXiv:1405.6635 [astro-ph.HE])

\bibitem{ASC} N. Anderson, T. Sidery and G. L. Comer, Superfluid neutron star turbulence, Mon. Not. R. Astr. Soc., {\bf 38} pp 1747-56 (2007). ({\it Preprint} arXiv:astro-ph/0703257)

\bibitem{SON} A. Sourie, M. Oertel and J. Novak, Numerical models for stationary superfluid neutron stars in general relativity with realistic equations of state, Phys. Rev. D {\bf 93} 083004 (2016). ({\it Preprint} arXiv:1602.06228 [astro-ph.HE])

\bibitem{CS} B. Carter and L. Samuelson, Relativistic mechanics of neutron superfluid in magneto elastic star crust, Class. Quant. Grav., {\bf 23} pp 5367-88 (2006). ({\it Preprint} arXiv:gr-qc/0605024)

\bibitem{CC} B. Carter and E. Chachoua, Newtonian mechanics of neutron superfluid in elastic star crust, Int.J.Mod.Phys. D {\bf 15} pp 1329-58 (2006). ({\it Preprint} arXiv:astro-ph/0601658)

\bibitem{PA} G. Pappas and T. A. Apostolatos, Revising the multipole moments of numerical spacetimes and its consequences, Phys. Rev. Lett., {\bf 108} 231104 (2012). ({\it Preprint} arXiv:1201.6067 [gr-qc])

 \bibitem{DWB} N. Degenaar, R. Wijnands, E. F. Brown, D. Altamirano, E. M. Cackett, J. Fridriksson, J. Homan, C. O. Heinke, J. M. Miller, D. Pooley, and G. R. Sivakoff, Continued neutron star crust cooling of the 11 Hz X-ray pulsar in Terzan 5: A challenge to heating and cooling models ?, Astrophys. J. {\bf 775}, 48 (2013). https://iopscience.iop.org/article/10.1088/0004-637X/775/1/48

\bibitem{TAP} A. Turlione, D. N. Aguilera, J. A. Pons, Quiescent thermal emission from neutron stars in low-mass X-ray binaries, Astron. Astrophys. {\bf 577}, A5 (2015). https://doi.org/10.1051/0004-6361/201322690

\bibitem{F8} J. A. Font, Living Rev. Relativity, {\it Numerical Hydrodynamics and Magnetohydrodynamics in General Relativity} (2008). http://www.livingreviews.org/lrr-2008-7.

\bibitem{NS} N. Stergioulas, Living Rev. Relativity, {\it Rotating Stars in Relativity} (2003). http://www.livingreviews.org/lrr-2003-3.

\bibitem{BC} B. Carter, {\it Relativistic superfluid models for rotating neutron stars, Trento, Italy, 2000} (Physics of the neutron star interiors, Eds. D. Blasche, N. K. Glendenning, A. Sedrakian) (astro-ph/0101257, 2001).

\bibitem{RS} M. A. Ruderman and P. G. Sutherland, {\it Rotating superfluid in neutron stars}, Astrophys. J. {\bf 190}, pp. 137-140 (1974).

\bibitem{NC} N. Chamel, {\it Two fluid models of superfluid neutron star cores}, Mon. Not. R. Astron. S. {\bf 388}, pp. 737-752 (2008).

\bibitem{AC2} N. Andersson and G. L. Comer, {\it Superfluid neutron star turbulence}, Mon. Not. R. Astron. S. {\bf 381} (2): 747 (2007).

\bibitem{FI} J. L. Friedman and J. R. Ipser, {\it Rapidly rotating relativistic stars}, Phylosophical Transactions: Physical Sciences and Engineering {\bf 340} No. 1658, Classical General Relativity pp. 391-422 (1992).

\bibitem{AWV} N. Andersson, S. Wellsand and J. A. Vickers, {\it Quantized vortices and mutual friction in relativistic superfluids}, Class. and Quantum Grav. {\bf 33} No. 24 (2016).

\bibitem{DL} D. Langlois, D. M. Sedrakian and B. Carter, {\it Differential rotation of relativistic superfluid in neutron stars}, arXiv:astro-ph/9711042, (1997).

\bibitem{291} M. L. Norman and K. -H. A. Winkler, {\it Why Ultrarelativistic Numerical Hydrodynamics is Difficult}, in M. L. Norman and K. -H. A. Winkler eds. {\it Astrophysical Radiation Hydrodynamics}, Proceedings of the NATO Advanced Research Workshop, Garching, Germany, August 2-13, 1982, NATO ASI Series C, vol. {\bf 188}, pp. 449-475, (Reidel, Dordrecht, Netherlands; Boston, U.S.A., 1986).

\bibitem{WE} S. Weinberg, {\it Gravitation and Cosmology} (John Wiley, 1972).

\bibitem{LPAP} L. Papantonopoulos {\it Physics of Black Holes: A Guided Tour (Lect. Notes Phys., Vol. 769),} (Springer, 2009).

\bibitem{MON} V. Moncrief, ``Gravitational perturbations of spherically symmetric systems: I. The exterior problem,'' Annals Phys. {\bf 88}, 323 (1974).

\bibitem{GP} A. Garat and R. Price, ``Gauge invariant formalism for second order perturbations of Schwarzschild spacetimes,'' Phys. Rev. D, {\bf 61}, 044006 (2000).

\bibitem{dsmg} A. Garat, Dynamical symmetry breaking in geometrodynamics {\it TMF} {\bf 195:2} 313-328 (2018).

\bibitem{dsmg2} A. Garat, Dynamical symmetry breaking in geometrodynamics {\it Theoret. and Math. Phys.} {\bf 195:2} 764-776 (2018) arXiv:gr-qc/1306.0602.





\end{thebibliography}

\end{document}